%% file: driverFile.tex
\documentclass{tcibook}

\usepackage{fancyhea}
\usepackage{work}
\usepackage{chngcntr}

\usepackage{bm}       
\usepackage{graphicx}
\usepackage{hyperref}      

\usepackage{amssymb}
\usepackage{graphicx}
\usepackage{array}

\input workshopsymbols.tex      

\counterwithout{figure}{chapter} 
\counterwithout{table}{chapter} 

\begin{document}

\def\bibname{References}

\bibliographystyle{utphys}  

\raggedbottom

\pagenumbering{roman}

\parindent=0pt
\parskip=8pt
\setlength{\evensidemargin}{0pt}
\setlength{\oddsidemargin}{0pt}
\setlength{\marginparsep}{0.0in}
\setlength{\marginparwidth}{0.0in}
\marginparpush=0pt


\pagenumbering{arabic}

\renewcommand{\chapname}{chap:intro_}
\renewcommand{\chapterdir}{.}
\renewcommand{\arraystretch}{1.25}
\addtolength{\arraycolsep}{-3pt}


\input NovelProbes/novel.tex



\end{document}

%% file: workshopsymbols.tex
\newcommand{\nc}{\newcommand}  

\def\beq{\begin{equation}}
\def\eeq#1{\label{#1}\end{equation}}
\def\eeqn{\end{equation}}

\newenvironment{Eqnarray}%
   {\arraycolsep 0.14em\begin{eqnarray}}{\end{eqnarray}}
\def\beqa{\begin{Eqnarray}}
\def\eeqa#1{\label{#1}\end{Eqnarray}}
\def\eeqan{\end{Eqnarray}}

\nc{\ra}{\rightarrow}  
\nc{\slsh}{\slash\hspace*{-0.22cm}}
\def\Re{{\cal R \mskip-4mu \lower.1ex \hbox{\it e}\,}}
\def\Im{{\cal I \mskip-5mu \lower.1ex \hbox{\it m}\,}}

\nc{\vev}[1]{ \left\langle {#1} \right\rangle }
\nc{\bra}[1]{ \langle {#1} | }
\nc{\ket}[1]{ | {#1} \rangle }
\nc{\fb}{\,{\rm fb}^{-1}}
\nc{\ev}{{\rm eV}}
\nc{\kev}{{\rm keV}}
\nc{\Mev}{{\rm MeV}}
\nc{\gev}{{\rm GeV}}
\nc{\tev}{{\rm TeV}}
\nc{\mev}{{\rm MeV}}

\def\del{\partial}
\def\Dslash{\not{\hbox{\kern-4pt $D$}}}
\def\dslash{\not{\hbox{\kern-2pt $\del$}}}
\def\pslash{\not{\hbox{\kern-2pt $p$}}}
\def\ETmiss{ \not{\hbox{\kern-4pt $E$}}_T }

\def\msb{{\bar{\ssstyle M \kern -1pt S}}}



%% file: NovelProbes/novel.tex
\newcommand{\nn}{\nonumber}
\newcommand{\f}{\frac}

\def\({\left(}
\def\){\right)}
\newcommand{\eotwash}{E\"ot-Wash}
\newcommand{\meff}{m_\mathrm{eff}}

\newcommand{\qb}{\mathbf{q}}
\newcommand{\pb}{\mathbf{p}}
\newcommand{\kb}{\mathbf{k}}
\newcommand{\xb}{\mathbf{x}}
\newcommand{\pd}{(2\pi)^d}
\newcommand{\intx}{\int d^d \mathbf{x}\ }
\newcommand{\rd}{{i8\rm d}}
\newcommand{\vp}{\varphi}
\def\gsim{ \lower .75ex \hbox{$\sim$} \llap{\raise .27ex \hbox{$>$}} }
\def\lsim{ \lower .75ex \hbox{$\sim$} \llap{\raise .27ex \hbox{$<$}} }

\renewcommand*\thesection{\arabic{section}}

\chapter*{Novel Probes of Gravity and Dark Energy}
\label{chap:novel-probes}

\begin{center}\begin{boldmath}

\input NovelProbes/authorlist.tex

\end{boldmath}\end{center}

\vspace{2truecm}
\thispagestyle{empty}



\section{Executive Summary}

The discovery of cosmic acceleration has stimulated theorists to consider dark energy or modifications to Einstein's General Relativity as possible explanations. The last decade has seen advances in  theories that go beyond smooth dark energy -- modified gravity and 
interactions of dark energy.
While the theoretical terrain is  being actively explored, the generic presence of fifth forces and dark sector couplings suggests  a set of distinct observational signatures. This report focuses on observations that differ from the conventional probes that map the expansion history or large-scale structure. Examples of such novel probes are: detection of scalar fields via lab experiments, tests of modified gravity using stars and galaxies in the nearby universe, comparison of lensing and dynamical masses of galaxies and clusters, and the measurements of fundamental constants at high redshift. The observational expertise involved is very broad as it spans laboratory experiments, high resolution astronomical imaging and spectroscopy and radio observations. In the coming decade, searches for these effects have the potential for discovering fundamental new physics. We discuss how the searches can be carried out using experiments that  are already under way or with modest adaptations of existing telescopes or planned experiments. The accompanying paper on the Growth of Cosmic Structure describes complementary tests of gravity with observations of large-scale structure. 



\section{Introduction}

If $\Lambda$CDM is not the correct description of our universe, we can safely conclude two things. One,  finding the correct theory will change our view of both fundamental physics and cosmology.  Two, a variety of experimental approaches that are currently outside mainstream cosmology can shed light on the new physics. The current program in cosmic frontiers is most focused on smooth dark energy uncoupled to matter.  It measures the distance-redshift relation and the growth of structure on large scales, typically at redshifts between 0 and 1.
The novel probes program discussed here is complementary to this, with sensitivity to couplings and to transitions in gravity between cosmic acceleration on Hubble scales and standard gravity on solar system scales.  These are fundamentally new windows on the dark sector and provide further handles on the dark energy mystery.

Alternate models, described in detail below, may broadly be described as including dark sector couplings, modified gravity or clustered dark energy --  these are related in many cases.
One possible reason for focussing on the narrow theory range of quintessence type models 
of smooth dark energy is that no generic observational signatures of alternate theories were  known to exist. However in the last decade, theoretical work on dark sector couplings and modified gravity theories has provided some useful guidelines for experiments. Based on these insights,  a number
of new tests have been carried out just in the last few years.
We will describe below the theoretical motivations, summarize
recent experimental constraints and outline an experimental program for the next decade.

\section{Modified Gravity (MG): Theory}
\parskip=5pt
\normalsize

The observed cosmic acceleration may be due to new physics within the gravitational sector itself. Einstein gravity is very well tested within the local universe {\it i.e.}, the solar system, but this leaves open the possibility that it is not a good description at the largest scales in the universe. This program of investigating consistent infrared modifications of Einstein gravity goes under the name of {\it modified gravity}. This program faces both theoretical and observational challenges. As we will see, modifications of General Relativity almost ubiquitously introduce new degrees of freedom in the gravitational sector, which are typically Lorentz scalars. If coupled to visible matter, these scalar fields will mediate a fifth force, at cosmological distances, astrophysical ones and within the solar system. However, gravity is exquisitely well tested within the solar system, so there must necessarily be some mechanism which hides these new fields from local observations---these degrees of freedom must be {\it screened}. In order to understand why these additional scalars must be screened, we need to make a connection between fields that appear in a lagrangian and our classical physics notions of force and potential. The derivation of the force between two objects due to the scalar will naturally present some mechanisms by which we can hide this force in dense environments, like our solar system.

\subsection{Scalar fields and fifth forces in modified gravity}
To begin, consider a general theory of a scalar field coupled to matter:\footnote{Throughout we will use the mostly plus metric convention {\it i.e.}, $\eta_{\mu\nu} = {\rm diag}(-1, 1, 1, 1)$.}
Such scalar fields arise generically in extensions of GR or the standard model of particle physics, e.g. in massive gravity theories, the extra degree(s) of freedom coming from the the graviton's mass are described by a scalar field.  Even a quintessence-like light scalar  
is expected to be coupled to matter at some level,
unless there is a symmetry that forbids such coupling.
The presence of such a coupling automatically implies a long range
scalar force. The Lagrangian for a coupled scalar field can be written as
\begin{equation}
{\cal L} = -\frac{1}{2}Z^{\mu\nu}(\phi, \partial\phi,\ldots)\partial_\mu\phi\partial_\nu\phi-V(\phi)+g(\phi)\rho~,
\end{equation}
here we have used the shorthand $Z^{\mu\nu}$ to indicate that there can be derivative self-interactions of the field. Now, in the presence of some sources, if we  expand the field about its background solution $\bar\phi$ as $\phi = \bar\phi+\vp$ to obtain the equation of motion for the fluctuations, it will be of the schematic form
\begin{equation}
Z(\bar\phi)\left(\ddot\vp-c_s^2(\bar\phi)\nabla^2\vp\right)+m^2(\bar\phi)\vp = g(\bar\phi)\rho
\label{generalscalareom}~,
\end{equation}
%
where $\bar\phi$ depends on background quantities such as the density and location of matter.
In  field theory, the potential between two localized static sources can be obtained and from it the force between two objects due to the presence of the $\vp$ field:
\begin{equation}
F(r) \sim \frac{g^2(\bar\phi)}{Z(\bar\phi)c_s^2(\bar\phi)}\frac{e^{-\frac{m(\bar\phi)}{\sqrt{Z(\bar\phi)}c_s(\bar\phi)}r}}{r^2}
\label{scalarforceeqn}
\end{equation}
Next we consider the impact of this fifth force on our observed universe. 
\subsection{Ways to suppress scalar force in the solar system: screening mechanisms}

For a light scalar $\phi$, and with the other parameters ${\cal O}(1)$, we see from Eqn.~\ref{scalarforceeqn} that its perturbation $\vp$ mediates a gravitational-strength long range force $F\sim 1/r^2$. The question we want to ask is {\it how can we make this force small, so that it is unobservable in the solar system?} Tests of gravity in the solar system strongly constrain the presence of  additional degrees of freedom, necessitating screening mechanisms for the light degrees of freedom invoked in modified gravity.
There are three broad possibilities.

1. The simplest thing we could imagine doing is to just set the coupling of $\vp$ to matter, $g$, to be very small. This leads to consideration of quintessence type models where the dark sector does not talk to visible matter. For our purposes, however, we will not  regard these types of models as  modified gravity. To some extent the distinction is arbitrary, but here we will consider theories where the additional degrees of freedom couple {\it non-minimally} to the metric to be genuine theories of modified gravity. A  more interesting thing to do than have $g$ be universally small, is to let the coupling to matter be {\it environmentally} small. This notion leads us fairly directly to models of screening of {\it symmetron} type.

2. Alternatively one can allow the mass of the field, $m(\bar\phi)$, to become large in regions of high density, this idea leads quite naturally to screening of {\it chameleon} type. All viable $f(R)$ gravity models use chameleon screening. 

3. Finally, we may imagine making the kinetic function $Z(\bar\phi)$ large environmentally---this leads us to screening of the {\it kinetic} type. This  mechanism, where second derivatives are important, is also known as {\it Vainshtein} screening. The DGP model and recent work on galileon and massive gravity theories rely on Vainshtein screening. 

In summary, chameleon screening increases the scalar effective mass at high densities, reducing the range of the fifth force.  Symmetron models undergo a symmetry-restoring phase transition at high densities, in which they decouple from matter.  The Vainshtein mechanism screens  through a non-canonical kinetic energy, effectively reducing their matter couplings.  
Chameleon fields screen in density-dependent as well as environment-dependent ways, known respectively as the chameleon and thin-shell effects.  In high-density regions such as the laboratory, a high chameleon effective mass means that large short-range fifth forces will have decayed away by the $\gsim 100~\mu$m distances accessible to experiments.  In low-density regions, an object with a sufficiently large gravitational potential will source the scalar field only through a thin outer shell of matter, reducing its effective fifth-force coupling by many orders of magnitude.
Each of these models has residual fifth forces which can be constrained in the laboratory.

It is well-known that scalar field theories have trouble with large quantum corrections, such as the hierarchy problem involving the Higgs in the Standard Model or the $\eta$ problem in cosmic inflation. 
Chameleon scalars escape laboratory constraints through large effective masses, but quantum corrections $\propto m_\mathrm{eff}^4$ to their potentials mean that excessively large masses spoil the predictivity of the theory~\cite{Upadhye:2012vh}.  Corrections from matter loops can also give large masses to symmetrons and chameleons with large couplings.  
Additionally, the interpretation of these theories in a Wilsonian framework opens up the possibility of ultraviolet-completion within string theory or some other fundamental theory~\cite{Hinterbichler:2010wu, Nastase:2013los}.
\subsection{Recent developments: massive gravity and {\it galileon} theories}

The history of massive gravity goes back all the way to 1939, when Fierz and Pauli first wrote down the quadratic action for a massive spin-2 particle~\cite{Fierz:1939ix}. The theory remained somewhat of a theoretical oddity until van Dam, Veltman and Zakharov \cite{vanDam:1970vg, Zakharov:1970cc} noticed around 1970 that the Fierz--Pauli theory does not reduce to linearized GR in the limit in which the mass of the graviton goes to zero; instead, in this limit, there is an additional scalar polarization mode which does not decouple and leads to an ${\cal O}(1)$ departure from GR predictions for the bending of light. This fact is known as the {\it vDVZ discontinuity}, because it indicates that the Fierz--Pauli theory does not have a smooth $m\to 0$ limit. This discrepancy was resolved shortly thereafter by Vainshtein~\cite{Vainshtein:1972sx}, who noted that completing the non-interacting Fierz--Pauli theory to an interacting non-linear theory makes the limit smooth. In an interacting theory, this additional scalar field becomes strongly self-interacting as the graviton mass is taken to zero, causing it to be screened near heavy sources; this is the {\it Vainshtein mechanism} we have discussed above. In~\cite{deRham:2010ik, deRham:2010kj}, de Rham, Gabadadze and Tolley (dRGT) constructed a theory that is free of a pathological sixth mode (the Boulware-Deser ghost; this was shown conclusively in~\cite{Hassan:2011hr}). Much work has gone into studying this massive gravity theory: for example studies of cosmological solutions~\cite{D'Amico:2011jj, Koyama:2011yg,Gumrukcuoglu:2011ew} and black holes~\cite{Berezhiani:2011mt} have been undertaken. Further, it has been realized that the dRGT construction opens the door to build theories with multiple interacting spin-2 particles~\cite{Hassan:2011zd, Hinterbichler:2012cn}. These preliminary investigations are encouraging, but there is still much work to be done to understand these theories.

Simultaneous with the development of these theories of a massive graviton, there has been a surge of interest in higher-derivative scalar field theories, most notably the {\it galileon} and its cousins. Galileons are scalar fields which have second order equations of motion and which, in their simplest incarnations, are invariant under a shift symmetry
\begin{equation}
\phi\longmapsto \phi+c+b_\mu x^\mu~,
\label{galsymm}
\end{equation}
which looks like a galilean boost, hence the name. These theories actually arose from studying braneworld modifications of gravity, in particular the Dvali--Gabadadze--Porrati (DGP) model \cite{Dvali:2000hr}, in which the cubic galileon term arises in the decoupling limit~\cite{Luty:2003vm, Nicolis:2004qq}. In \cite{Nicolis:2008in}, the properties of the cubic galileon were abstracted to write down the most general action with the symmetry of Eq.~\ref{galsymm} with second order equations of motion. Remarkably, there are only 5 terms with these combined properties in 4d. They have remarkable field-theoretic properties; for example, they exhibit Vainshtein screening around heavy sources~\cite{Deffayet:2001uk}, enjoy a non-renormalization theorem~\cite{Luty:2003vm,Hinterbichler:2010xn} and can be interpreted as Wess--Zumino terms for the breaking of space-time symmetries~\cite{Goon:2012dy}. Since the introduction of the additional galileon model, they have been generalized in myriad ways and used for many applications, ranging from the early to the late universe. Although the galileons were first investigated in a braneworld context, it turns out that they are also intimately connected to massive gravity, appearing as the scalar mode of the massive graviton in the decoupling limit, \cite{deRham:2010ik,deRham:2012az}. The coupling to matter might be slightly different in the case of massive gravity, leading to potential non-conformal couplings which could affect lensing,~\cite{Wyman:2011mp}.

 \section{Observable consequences of gravity theories}
\begin{figure}[tb]
\begin{center}
\includegraphics[width=5.5in]{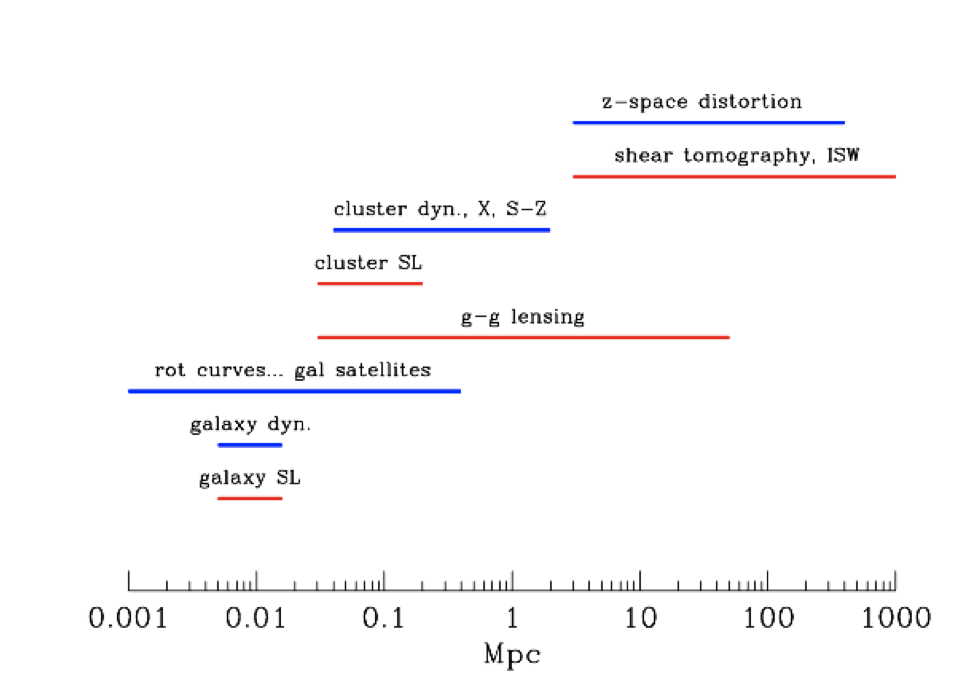}
\caption{Tests of gravity at different length scales. Red lines shows observations that probe the sum of metric potentials $\Phi + \Psi$ typically via gravitational lensing, while blue lines show dynamical measurements that rely on the non-relativistic motions of stars or galaxies and are sensitive to  $\Psi$ alone. Adapted from the review by Jain \& Khoury \cite{JainKhoury}. 
\label{gravityscales}
}
\end{center}
\end{figure}

A  general way to see the observable signatures of MG theories is to consider 
he perturbed Friedman-Robertson Walker metric for scalar tensor theories: 
\begin{equation}
ds^2  = -(1+2\Psi) dt^2 + (1-2\Phi)\ a(t)^2 d\vec{x}^2 
\end{equation}
The two metric potentials $\Phi$ and $\Psi$ are identical for quintessence-like GR cosmologies, 
and indeed for any dark energy model that is based on a novel type of matter or fluid.
The two potentials are, however, generically 
not equal in MG.
This leads to a difference from GR when one considers the 
motions of non-relativistic tracers that respond to the Newtonian potential $\Psi$, as shown 
explicitly in the force law of Eqn.~\ref{scalarforceeqn} above. 
Two of the most important signatures of MG theories follow:

\begin{itemize}
\item MG theories of interest behave as scalar-tensor theories on observable scales.
Such theories generically break the equivalence between the mass
  distribution inferred from the motions of stars and galaxies
  (non-relativistic tracers) versus that inferred from
  photons. Thus the comparison of dynamical and lensing masses of
  galaxies, clusters and large-scale structures can yield signatures
  of MG. 
\item MG theories rely on screening mechanisms that shield high
  density regions like the halo of the Milky Way from modified
  forces. The screening must ensure that stringent solar system and
  lab tests of GR are satisfied. Hence there is a transition
  from gravity being GR-like within large halos to
  being modified on large scales and/or for smaller halos.
  Potentially observable deviations  in the
  dynamics of stars and galaxies arise from the fifth (scalar) forces in this regime.
\end{itemize}

Screening mechanisms typically utilize some measure of the mass distribution
 of halos, such as the density or  Newtonian potential, to recover GR well within the Milky Way.
This leaves open the possibility that smaller halos, the outer parts of  halos, or some components of the mass distribution, can experience enhanced forces. For a given mass distribution,
unscreened halos will then have higher internal velocities and center of mass
velocity compared to GR. This can produce deviations of $\sim$10-100\%  from GR, with
distinct variations between different mechanisms in the size of the effect and the
way the transition to GR occurs. It is important to note that observable effects are typically larger on
halo scales than in the linear regime or at high redshift.
Since MG models recover GR at high redshift for consistency with CMB and Nucleosynthesis observations, effects of enhanced forces are manifested only at late times. Both these facts favor tests
in the nearby universe. At the very least, tests in nearby stars and galaxies offer a complementary probe of gravity theories.  See
the Growth of Cosmic Structure paper \cite{growth}  for more discussion of cosmological tests. 

In what follows we will treat chameleon, symmetron and environmentally dependent dilaton screening mechanisms in one category as their qualitative observational signatures are similar. The second category is Vainshtein theories whose signatures are distinct from
chameleon-type theories.  The tests we describe below will contain two fundamental parameters of the theories: the coupling of the  fifth force to matter and the range of interaction of the fifth force. The meaning of the second  parameter is model dependent. 

An overview of the new observational tests of gravity is provided in Figures \ref{gravityscales}, \ref{chameleonlimits} and \ref{figure:tests-observations} and the recommendations are summarized in Tables \ref{gravitytable} and \ref{futuretable}. 

\subsection{Lensing and Dynamical Masses}
The deflection law for photons that leads to various gravitational lensing effects is the same
in any metric theory of gravity. Moreover the relation of light deflection to the mass distribution
is largely unaltered in scalar-tensor theories; lensing masses are true masses. However
the acceleration of galaxies, which move at non-relativistic speeds, is altered as the Newtonian
potential is different from that in GR: it receives additional contributions from the scalar field in
scalar-tensor models. Thus a discrepancy arises in the mass distribution inferred from
dynamical tracers and lensing.

A comparison of lensing and dynamical cross-power spectra was proposed by~\cite{Zhang:2007nk} as a model independent test of gravity. This test is
in principle immune to galaxy bias, at least to first order, and can also
overcome the limitation of sample variance on large scales. Thus it can
be applied in the linear regime
relatively easily, provided both multi-color imaging (for lensing) and
spectroscopy (for dynamics) are available for the same sample of galaxies. Galaxy cluster surveys could  provide a and complementary comparison of dynamical versus lensing masses, through comparison of the kinetic Sunyaev-Zeldovich (SZ)  signal \cite{Kosowsky:2009nc,Keisler:2012eg}, the Doppler shift of the SZ signal due to the cluster's motion along the line of sight, with  lensing measurements. For a more detailed discussion of the redshift space measurement of 
of galaxy clustering, see the report on the Growth of Cosmic Structure\cite{growth}. 

\begin{figure}[tb]
\begin{center}
\includegraphics[width=4.5in]{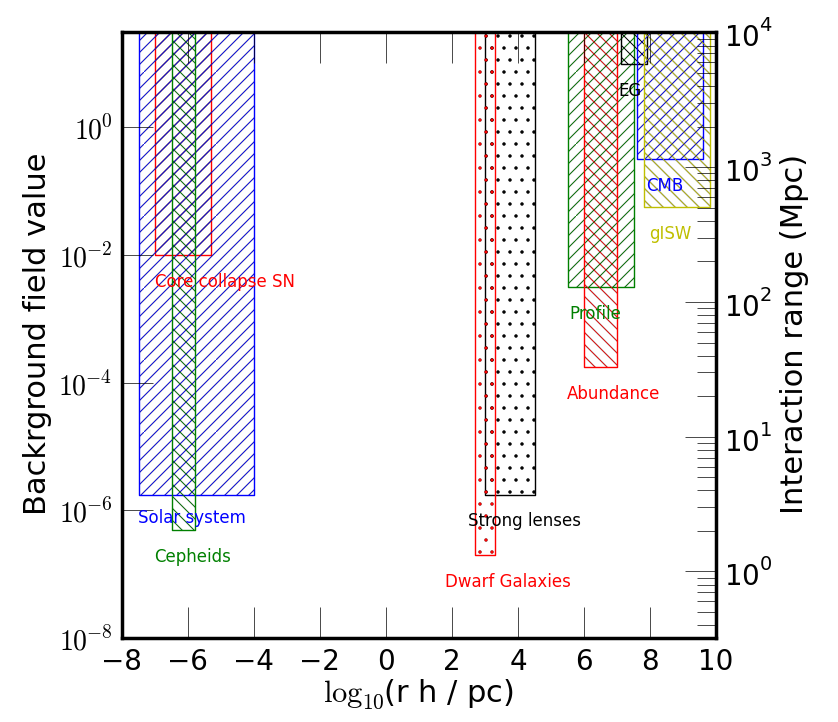}
\caption{Astrophysical~\cite{Hu:2007nk,Jain:2012tn,Vikram:2013uba} and cosmological~\cite{Song:2007,Giannantonio:2010,Schmidt:2009} limits on chameleon theories. The spatial scale on the x-axis gives the range of length scales probed by particular experiments. The parameter on the y-axis is the background field value, or the range of the interaction (y-axis label on the right side) for an $f(R)$ model of the accelerating universe. The rectangular regions give the exclusion zone from a particular experiment. All  but the solar system results have  been obtained in the last 5 years, illustrating the impressive interplay between theory and experiment in the field. The two rectangles with dots are meant to indicate preliminary results from ongoing work. This figure is adapted from Lombriser et al \cite{Lombriser:2012}. 
\label{chameleonlimits}
}
\end{center}
\end{figure}

On much smaller scales, versions of the test
can be performed by comparing  lensing and dynamical masses of galaxies
and clusters.
This test is fairly unique to testing gravity, as it has
little information to add in the dark energy framework.
At least three kinds of tests are available:
the comparison of strong lensing with measured stellar
velocity dispersions in the inner parts of elliptical galaxies~\cite{Bolton:2005nf},
the virial masses of halos from weak lensing and dynamics~\cite{Schmidt:2010jr},
and the infall region that extends to ten or more times the virial radius~\cite{Reyes:2010tr}.
The latter two tests are feasible only for massive clusters or using stacked
measurements of large samples of galaxies binned in luminosity or another observable
that serves as a proxy for halo mass.  Figures \ref{gravityscales} and \ref{chameleonlimits} 
include the tests described above as part of a wider set of tests of gravity. 

\subsection{Astrophysical tests in the nearby universe}
\label{section:astrotests}

The logic of screening in scalar-tensor
gravity theories implies that on small scales the fifth force that arises in MG
impacts some tracers and not others.  Galaxies themselves can have enhanced motions, as discussed above in the tests involving lensing and dynamical masses. 
And the components of galaxies --
stars, gas, neutron stars and black holes -- respond differently to the fifth force because
they can be screened at different levels. The different components can acquire different velocities or be displaced in their spatial distribution. A variety of observable phenomena result, and are typically best observed in nearby galaxies which can be seen at high resolution~\cite{Jain:2011ji}. 
We will continue to classify theories into two classes based on the screening mechanism used 
to recover GR in the solar system: chameleon and Vainshtein theories. 

For chameleon theories
galaxies in low-density environments may remain unscreened as the
Newtonian potential $\Psi_{\rm N}$, which determines the level of screening,
can be smaller than in the Milky Way. Hence
dwarf galaxies can exhibit manifestations of modified forces in both
their infall motions and internal dynamics. For Vainshtein theories, the velocities of
galaxies and other tracers of gravity are enhanced (and can compared to lensing,
as described above) and compact objects can separate from stars and gas. The magnitude of these observable effects has been estimated for $f(R)$ and DGP-like theories. We first 
describe the effects qualitatively and then summarize the expected magnitude at the end 
of this subsection. 

Stars within unscreened galaxies may show the effects of modified gravity.
In \cite{Chang:2010xh} and~\cite{Davis:2011qf} the effects on  giant
and main sequence stars, respectively are described: essentially the enhanced gravitational force
makes stars of a given mass brighter and hotter than in GR. They are also more ephemeral since they consume their fuel at a faster rate.
For the Sun the  potential  $\Psi_{\rm N} \approx 2\times 10^{-6}$ ;  coincidentally, the potential of the Milky Way is close to this value, and is
believed to be sufficient to screen the galaxy so that solar system tests of
gravity are satisfied.
 Thus main sequence stars whose masses are similar to that of the Sun
are likely to be partially or completely screened.
Red giants are at least ten times larger in size than the main
sequence star from which
they originated so they have $\Psi_{\rm N} \sim 10^{-7}$ -- thus their envelopes
may be unscreened. The enhanced forces lead to smaller radii, hotter
surface temperatures and higher luminosities than their GR doppelg\"angers.

Specific stages of the evolution of  giants and supergiants are used to obtain
accurate distance estimates; they also happen to provide useful tests of gravity.
 The first feature, well known
for its use in obtaining distances, is the nearly universal
luminosity of  $\lsim 2 M_\odot$ stars at the tip of the red giant branch (TRGB).
The second is the
Period-Luminosity
relation of cepheids, which are giant stars with $\sim 3-10 M_\odot$ that
pulsate when their
evolutionary tracks cross a near universal, narrow range in temperature known as the
\textit{instability strip}.
The tight relation between luminosity and other observables is what makes these stars valuable distance indicators -- it also
makes them useful for tests of gravity.
For background field values  in the range $10^{-6}$--$10^{-7}$, the
 TRGB luminosity is largely robust to modified gravity while the cepheid period-luminosity
relation is altered. Measurements of these properties
within screened and unscreened galaxies then provide tests of gravity:
the two distance indicators should agree for screened galaxies but not for
unscreened galaxies~\cite{Jain:2012tn}.

In disk galaxies, stars and gas can respond differently to the fifth force in chameleon theories.
For unscreened dwarf galaxies,
the rotation of the stellar disk can be slower than that of the HI gas disk because the stars are screened for some range of theory parameters. The rotation of the stellar and gas disk 
can be measured using different optical and radio observations respectively.
Further, the external fifth force on a dwarf galaxy can cause
the stars, giant stars and gas to segregate along the direction of the external force. For small
disk galaxies, this also leads to a warping of the stellar disk~\cite{Vikram:2013uba}. In Vainshtein theories compact
objects, especially black holes, are immune to the scalar fifth force owing to a no-hair theorem. 
So the central, supermassive black hole lags the stars and gas as the galaxy moves in response to
external forces~\cite{Hui:2012jb}.  All these effects are potentially
observable; preliminary tests of gravity using data in the literature have been carried out
or are ongoing. Figure \ref{chameleonlimits} shows some of the upper limits obtained from
these tests.

In more massive galaxies and clusters, the infall of satellites and dark matter is enhanced
due to the fifth force. Specific signatures in the density profiles of galaxies are detectable
using lensing measurements. Comparison of lensing and redshift space observations provide
tests of Vainshtein theories as well.

A varied array of other probes offers the possibility of testing chameleon theories astrophysically. Observations of circularly polarized starlight in the wavelength range $1-10^3\AA$ could be a strong indication of chameleon-photon mixing \cite{Burrage:2008ii}. A variation in the ratio of the electron to proton mass between the laboratory and in space can be measured and indicate the existence of a chameleon like scalar particle at work. \cite{Levshakov:2010cw}. As demonstrated by  \cite{Burrage:2008ii,Levshakov:2010cw}
the scatter in the luminosity of astrophysical objects can be used to search for chameleons in particular through the observations of active galactic nuclei.

To summarize,
screening mechanisms yield distinct signatures in
many settings; a selective list follows:  
\begin{itemize}
\item Enhanced velocities of unscreened galaxies. In both Vainshtein and Chameleon theories 
enhancements of order ten percent, or tens of km/s  are expected (given that typical peculiar velocities are a few hundred km/s). 
\item Segregation of the different components of galaxies. In Chameleon theories the gas and stellar components respond differently to the external forces, an example is that the stellar disk may be warped relative to the neutral H disk. In Vainshtein theories the supermassive black hole may be displaced by up to 0.1 kpc from the stellar light, along the direction of the external force. 
\item Differential internal dynamics of different tracers. The stellar disk may rotate more slowly than the gas disk in Chameleon theories. Systematic differences of 5-10 km/s  are expected in dwarf galaxies with circular velocities below 100 km/s. 
\item Stellar evolution is altered, with giant stars in particular moving more rapidly through their evolutionary tracks in Chameleon theories. The most distinct observable consequences are for 
distance indicators, in particular the comparison of cepheid variables and TRGB stars. Relative 
offsets at the 5\% level  in the distance ladder are expected. 
\end{itemize}
The approaches to detecting the effects listed above and distinguishing them from astrophysical sources have been discussed in the literature cited above.  Figures \ref{gravityscales} and \ref{chameleonlimits} show the regimes and constraints from some of these tests. 
 One key element to a convincing detection is
the creation of screened (control) and unscreened samples of galaxies or 
other tracers. Further correlating the observed effect with the direction of the scalar fifth force rules out the primary sources of astrophysical uncertainty. 
 The current status and future prospects of these tests are discussed below and summarized in Table \ref{gravitytable}.

\subsection{Cosmological Probes}

Large-scale probes of gravity may be classified as follows: 1. Tests of the consistency of
the expansion history and growth of structure. A discrepancy in the inferred $w$ from the
two approaches can signal a breakdown of the GR-based smooth dark energy cosmological
model. 2. Detailed measurements of the linear growth factor at different scales and redshift. 3. A
comparison of the mass distribution inferred from different probes, in particular redshift
space distortions and lensing. The latter is a compelling test of MG, as described above, 
since the same test can be carried out over many scales. Table \ref{gravitytable} summarizes
the current state and future prospects for the different tests. 

As with tests of dark energy, 
the linear regime offers the twin advantages of ease of prediction and interpretation 
together with immunity from astrophysical systematics which typically cannot
alter structure formation on scales above 100 Mpc. And specific MG models can
produce scale and redshift dependent growth that in principle distinguishes them
from dark energy models with the same expansion history. The main limitation
for tests of gravity is that the signal is typically smaller on large scales and
in practice is degenerate with effects such as scale dependent galaxy bias. See
the report on the Growth of Cosmic Structure \cite{growth} for a discussion. An important area for future work 
is to make explicit the connection between large-scale tests and the small scales ones
discussed in this chapter. In particular, we may be faced with compelling questions like: 
if new physics is detected using astrophysical tests, what is the implication for cosmological theories and large-scale tests? 

\subsection{Laboratory tests of modified gravity and dark sector interactions}
\label{s:labtests}

Although laboratory experiments cannot directly probe constant or slowly-varying dark energy potentials, they provide the most sensitive constraints on certain classes of interactions between dark energy and Standard Model particles.  Gravitation-strength fifth forces are best probed using torsion pendulum experiments in the laboratory.  More strongly coupled models, in which screening is more powerful, can be probed using cold neutron systems.  Dark energy models which couple to photons as well as to matter can be produced and trapped in afterglow experiments, or produced in the Sun and detected in magnetic helioscopes.  
For this section we adopt the effective potential $V_\mathrm{eff}(\phi) = V(\phi) + \beta \phi T_\mu^\mu/M_\mathrm{Pl} + \frac{1}{4}\beta_\gamma \phi F_{\mu\nu}F^{\mu\nu}$, where $T_{\mu\nu}$ in the matter interaction is the stress-energy tensor and $F_{\mu\nu}$ in the photon interaction is the electromagnetic field strength.
The tests proposed below are only what has been considered so far in a relatively young but growing subfield. Laboratory tests of dark energy apply techniques from the Intensity Frontier to constrain Cosmic Frontier physics.

\begin{figure}[th]
\begin{center}
\includegraphics[width=5.1in]{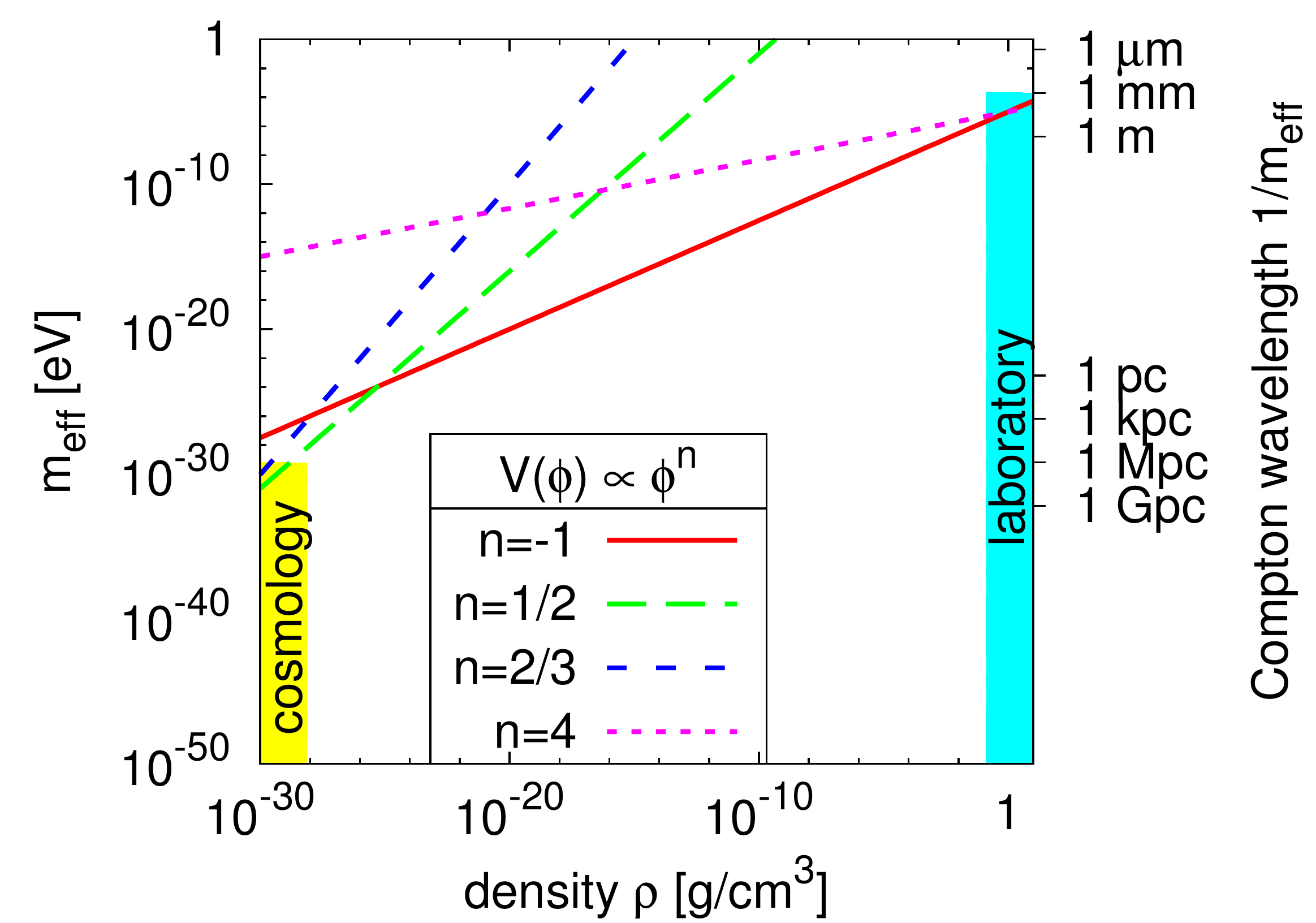}
\caption{Mass scaling in chameleon models with $V(\phi) = \textrm{const} + g |\phi|^n$.  Cosmological probes exclude gravitation-strength fifth forces with ranges $\gsim 1$~Mpc, while laboratory experiments exclude such forces with ranges $\gsim 1$~mm at much higher densities -- shown by the shaded regions.  Thus laboratory constraints will be dominant for models  $n \lesssim -1/2$ and $n>2$ with less steep mass scalings, while cosmological constraints will dominate for $-1/2 \lesssim n < 1$. 
\label{f:which_scale}}
\end{center}
\end{figure}

{\bf Fifth force tests:}  Although screening can reduce the fifth force range or effective matter coupling in a scalar field dark energy candidate, laboratory experiments provide powerful constraints on the residual forces~\cite{Adelberger:2003zx,Adelberger:2009zz}.
Gravitation-strength fifth forces due to a scalar with mass $\meff$ are best tested by measuring the force between two objects of size $\sim 1/\meff$ separated by distances $\sim 1/\meff$.  Chameleon models with matter coupling strength $\beta \sim 1$ and well-controlled quantum corrections (``quantum-stable'' models) tend to have $\meff \lsim 0.01$~eV at laboratory densities, making \eotwash~an ideal experiment~\cite{Upadhye:2012vh,Kapner:2006si,Adelberger:2006dh,Upadhye:2012qu}.
\eotwash~uses a torsion pendulum in which two parallel disks have a matched set of surface features.  As the lower disk is uniformly rotated, gravity and any fifth forces torque the upper disk so as to align the two.
Current \eotwash~bounds exclude a range of chameleon potentials, as well as symmetron models with masses around the dark energy scale and coupling energies just beyond the Standard Model scale.
As shown in Figure~\ref{f:which_scale}, laboratory constraints on chameleon models are complementary to cosmological constraints. Models whose effective masses scale rapidly with density, such as $n=1/2$ in the figure, have vanishing fifth forces in the laboratory and are best probed at low densities.  Models with $n>2$ or $n\lsim -1/2$ can best be constrained by looking at the smallest possible distances, which for gravitation-strength fifth forces means sub-millimeter torsion pendulum experiments such as \eotwash.  
Since such laboratory-detectable chameleons typically have Compton wavelengths at the dark energy scale $\sim 100~\mu$m, they are insensitive to the gravitational potentials of the Earth, Sun, and Galaxy.

Dark energy models which couple more strongly to matter are too well screened to be detected by torsion pendulum experiments, but are still accessible to other types of tests~\cite{Mota:2006ed,Mota:2006fz}.
Fifth forces can be detected in Casimir force experiments~\cite{Brax:2007vm}.  Since the Casimir force is an electromagnetic effect much stronger than the force of gravity, Casimir experiments probe fifth forces with much stronger couplings.  Current experiments
exclude quantum-stable quartic chameleons with couplings $\beta \gsim 10^6$.  More recently, cold neutron experiments have been used to constrain dark energy. Bouncing neutron experiments, which measure the $\sim 1$~peV energy splittings of a neutron in Earth's gravitational field, also probe fifth forces between the neutrons and the experimental apparatus~\cite{Brax:2011hb,Pokotilovski:2013gta}.  Since the current Grenoble experiment measures these splittings at the $\sim 10\%$ level, it constraints fifth forces from the apparatus which are nearly as large as the gravitational force of the entire Earth.  Thus Grenoble excludes very large matter couplings, $\beta \gsim 10^{11}$.  These experiments are summarized in Table~\ref{t:expts}.

{\bf Searches for photon couplings:}  If dark energy couples to photons as well as to matter, then dark energy particles can be produced through photon oscillations in a background magnetic field, in much the same way as axions~\cite{Brax:2007hi}.  Matter interactions can be used to trap these dark energy particles inside a vacuum chamber.  An ``afterglow experiment'' turns off the photon source after a population of dark energy particles has built up, and looks for photons regenerated through oscillations~\cite{Ahlers:2007st,Gies:2007su,Chou:2008gr}.  Trapping of dark energy particles requires that they be massive in the chamber walls, and oscillation requires that they be light in the interior of the chamber.   GammeV at Fermilab was the first afterglow experiment, constrained chameleon dark energy by making a simple configuration change to an axion experiment~\cite{Chou:2008gr}.  Its successor CHASE has extended its constraints by orders of magnitude~\cite{Steffen:2010ze,Upadhye:2009iv,Upadhye:2012ar}, and further afterglow experiments have been proposed at JLab and Tore Supra. The design of the CHASE experiment is such that the photon coupling is constrained subject to the matter coupling being within a certain range. However, CHASE does not constrain the matter coupling $\beta$ itself. If a signal is seen and the particle mass is constrained, more sensitive experiments using microwave cavities can confirm the detection.  The ADMX group has conducted an afterglow experiment in a microwave cavity~\cite{Rybka:2010ah}, and the Yale Microwave Cavity Experiment group is planning a similar cavity experiment~\cite{Baker:2013zta}.  Photon-coupled dark energy particles could also be produced in the Sun and reconverted into photons in a magnetic helioscope.  Since a helioscope does not need to trap a particle in order to detect it, helioscopes can detect or exclude particles which are too light to trap in an afterglow experiment.  Simple changes to the configuration of the CAST axion helioscope~\cite{Zioutas:2004hi} would allow it to search for photon-coupled dark energy.  With remarkably low cost we have the potential of great impact.

\begin{table*}[th]
\begin{center}
\begin{tabular}{|l||c|c|}
\hline
Experiment &   Type           & Couplings excluded                \\
\hline
\hline
\eotwash   & torsion pendulum & $0.01 \lsim \beta \lsim 10$ \\
\hline
Lamoreaux  & Casimir          & $\beta \gsim 10^5$ ($\phi^4$)   \\
\hline
Grenoble   & bouncing neutron & $\beta \gsim 10^{11}$           \\
\hline
GRANIT     & bouncing neutron & forecast: $\beta \gsim 10^8$    \\
\hline
NIST       & neutron interferometry & forecast: $\beta \gsim 10^7$ \\
\hline
CHASE      & afterglow        &                                $10^{11} \lsim \beta_\gamma \lsim 10^{16}$ subject to \mbox{$10^4 \lsim \beta \lsim 10^{13}$,} \\
\hline
ADMX       & microwave cavity & \mbox{$\meff = 1.952~\mu$eV,}                                                   $10^9 \lsim \beta_\gamma \lsim 10^{15}$\\
\hline
CAST       & helioscope       & \mbox{forecast: $\beta \lsim 10^{9}$,}                                       $\beta_\gamma > 10^{10}$\\
\hline
\end{tabular}
\end{center}
\caption{Laboratory tests of dark energy interactions with matter ($-\beta T_\mu^\mu / M_\mathrm{Pl}$) and electromagnetism ($\frac{1}{4} \beta_\gamma \phi F_{\mu\nu}F^{\mu\nu} / M_\mathrm{Pl}$).  Constraints apply to chameleon models with potential $V(\phi) = M_\Lambda^4(1 + M_\Lambda/\phi)$ and $M_\Lambda = 2.4\times 10^{-3}$~eV (unless otherwise noted).  \label{t:expts}}
\end{table*}

\subsection{Solar system tests of gravity}

Solar system measurements have historically been powerful tests of gravity.  Precise experimental confirmations of Kepler's Law impose stringent limits on deviations from the gravitational inverse-square law, which are predicted by unscreened modified gravities such as the Brans-Dicke theory.  However, solar system constraints on screened models are considerably weaker, and are not currently competitive with laboratory and cosmological constraints.

Chameleon models evade solar system constraints in two distinct ways. Figure~\ref{f:which_scale} divides chameleons into two subsets: ``laboratory'' chameleons, with $n \lsim -1/2$ or $n>2$, which are best probed in the lab; and ``cosmological'' chameleons, with $-1/2 \lsim n < 1$, roughly corresponding to inverse curvature terms in $f(R)$, which are constrained cosmologically.  
Since torsion pendulum experiments~\cite{Adelberger:2006dh} require that a laboratory chameleon have a Compton wavelength less than a millimeter at laboratory densities $\sim 1$~g/cm$^3$, each planet in the solar system would have a thin shell even in isolation.
The thin shell effect reduces the effective chameleon coupling by many orders of magnitude.  The Earth's coupling, for example, is reduced by $\sim (1\textrm{ mm} / 6400 \textrm{ km})^2 \sim 10^{-20}$, with similar results for the other planets.  
Thus solar system measurements are subdominant to other constraints on such models~\cite{Gubser:2004uf}.
Screening of $-1/2 \lsim n < 1$ (``cosmological'')
 chameleons, on the other hand, requires a large gravitational potential well.  Kepler's Law tests require that the Sun be screened.  If the solar system were isolated, then these constraints would imply that the chameleon self-screening parameter be smaller than the Sun's gravitational potential, $\approx 2 \times 10^{-6}$.  However, the solar system sits inside the Galaxy, itself a potential well of magnitude $\approx 2 \times 10^{-6}$, which in turn sits inside the Local Group and the Virgo Supercluster.  Environmental screening likely weakens this solar system constraint by a factor of a few~\cite{Hu:2007nk}.  Thus solar system constraints are an order of magnitude less constraining than those from Cepheid variables and dwarf galaxies.

For a subset of Galileon fifth forces, Refs.~\cite{Lue:2002sw,Battat:2008bu} prove the beautiful result that the angular precession of a planetary orbit is $3 c / (8 r_c) = 25 (1 \textrm{ Gpc}/r_c)~\mu$as/yr, independent of the mass and orbital radius of the planet.  
Slight differences in the planets' orbital inclinations allow such precessions to be measured, providing a bound $r_c > 125$~Mpc on the DGP model. Currently this bound is subdominant to cosmological constraints by over an order of magnitude.  However, substantial improvements to measurements such as lunar laser ranging are expected in the near future, which have the potential to make solar system constraints on DGP competitive.

In the DGP model, the situation is somehow similar for pulsar tests of General Relativity. The very presence of a field that acts as a Lorentz scalar opens up new channels of gravitational radiation in binary systems. Monopole and dipole radiation are in principle possible in the DGP model but these are very suppressed by both relativistic corrections and the Vainshtein mechanism which makes such tests uncompetitive compared to solar system tests let alone cosmological ones,~\cite{deRham:2012fw,Chu:2012kz}. However the situation might be different for more general models that exhibit a Vainshtein mechanism such as more general Galileon models. In these models any breaking of the spherical symmetry could lead to potential large corrections to the solutions and thus to the emitted radiation. This remains to be explored in more depth -- both at the theoretical  and  observational level ~\cite{deRham:2012fg}.

\section{Fundamental constants and coupled scalar fields}

Many novel cosmological scenarios, such as those than invoke rolling scalar fields, and beyond standard model physics, such as supersymmetry (SUSY), allow for the possibility that the fundamental constants have different values in the early universe than they have now. The study of fundamental constants is particularly relevant because it is able to probe both cosmology and physics beyond the standard model in tandem.
Modified gravity models couple to matter, and also possibly to the gauge fields.  An alternative is a dark energy which couples to gauge fields but not to matter, eliminating the need for a location-dependent screening mechanism.  Homogeneous evolution of dark energy in such models implies a time-variation of dimensionless quantities including the fine structure constant $\alpha$ and $\mu = m_p/m_e$, which may be measured with exquisite precision.
The fundamental constants measure the product of a cosmological quantity---$1+w$, the deviation of the dark energy equation of state from $-1$, a measure of how far the field rolls ---and a new physics quantity---the coupling of a rolling scalar field to the electromagnetic field, $\zeta_x$, where $x$ is either $\mu$  or $\alpha$.

If we assume $\Lambda$CDM cosmology and the standard model of particle physics, both  $\zeta_x^2$ and $(w+1)$ are zero. However, in cosmologies that invoke a rolling scalar field $(w+1)$ is non-zero and in beyond the standard model physics, such as supersymmetry, $\zeta$ is non-zero. Therefore, measuring either $\alpha$ or $\mu$ in the early universe tests for both new cosmology and new particle physics.

\subsection{Observational Method}

Tests of the constancy of the fundamental constants are carried
 out both in laboratories and astronomical observatories.
  Laboratory tests determine the values of the fundamental
   constants to high accuracy but have very short time bases to
    test for a possible variance.  Conversely, astronomical observations have extremely long time bases but much lower accuracy and no control over the environment where the measurements are made.  Measurements of the fine structure constant $\alpha$ are done, as the name implies, by looking at the fine structure transitions in atoms, either in the laboratory or in intergalactic gas by astronomical observations.  In the laboratory atomic clocks are used to make the measurement.  Astronomical measurements utilize spectrometers with various calibration standards.
    Measurements of the proton to electron mass ratio $\mu$ utilizes molecular transitions. In practice optical spectroscopy of redshifted Lyman and Werner absorption bands of $H_2$ in high redshift Damped Lyman
Alpha systems (DLAs) and radio observations of rotational, inversion and hindered transitions at lower redshift have provided the bulk of the astronomical data.  At present the astronomical observations have provided the most stringent bounds on the variance of a fundamental constant. In \cite{Bagdonaite} the change in $\mu$ is constrained to be $\delta \mu/\mu<10^{-7}$  at a redshift of 0.89.  This translates for a linear change in time to  $\dot{\mu}/\mu<1.4 \times 10^{-17}$   per year as compared to the laboratory limit of  $\dot{\mu}/\mu<5.6 \times 10^{-14}$ per year~\cite{Uzan:2010pm}.

\subsection{Fundamental Constants with Dark Energy and SUSY}
Commonly, such as in models of quintessence~\cite{Ratra:1987rm, Caldwell:1997ii} or $K$-essence~\cite{ArmendarizPicon:2000dh, ArmendarizPicon:2000ah}, cosmologies invoke a rolling scalar field, $\phi$, which couples to gravity to drive the accelerated expansion of the universe.  At the same time, new physics beyond the standard model such as supersymmetry also predicts variation of the fundamental constants.  The variation is proportional to the difference between the rate of change of the Quantum Chromodynamic Scale, $\Lambda_{\rm QCD}$, and the rate of change of the Higgs vacuum expectation value $v$.   Most treatments result in the change in $\mu$ being about 40 to 50 times larger than the change in $\alpha$~\cite{Avelino:2006gc}.

\begin{figure}
\begin{center}
\includegraphics[width=4.1in]{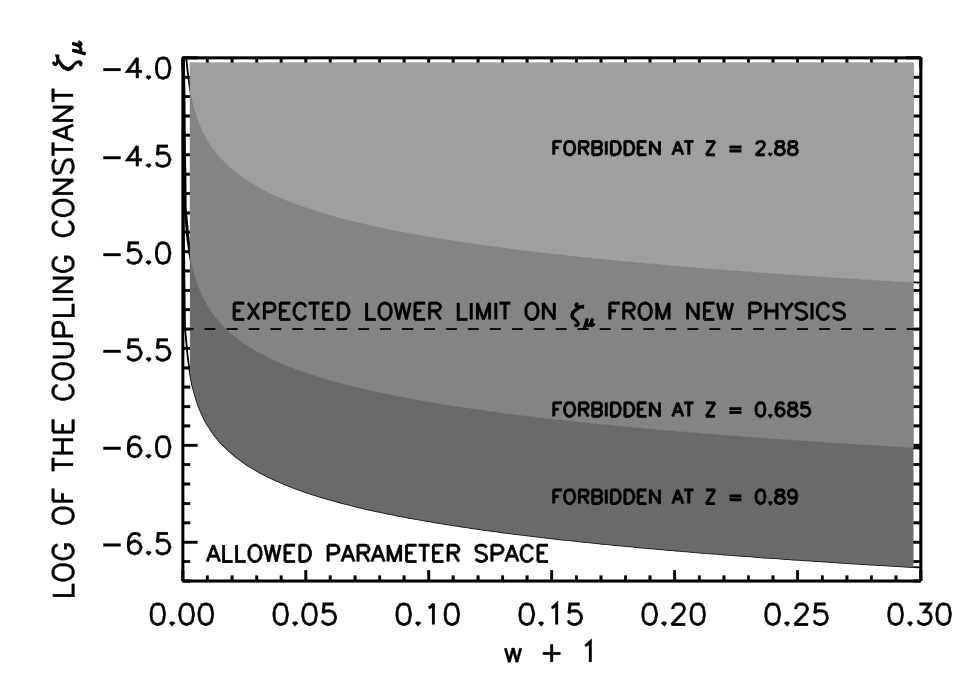}
\caption{ The current bounds on the dark energy equation of state and the coupling constant $\zeta_\mu$ from observational constraints on the variance of $\mu$.  At the labeled redshifts all of the area above the lower boundary of the shaded area is forbidden.  For example at a redshift of 0.89 all of the shaded area is forbidden.  At a redshift of 2.58 only the lightest shaded area is forbidden.}
\label{constants1}
\end{center}
\end{figure}

Following~\cite{Nunes:2003ff,Thompson:2012my}, we can relate the rolling scalar field, the dark energy equation of state and the fundamental constants.  Couplings between the scalar field $\phi$ and the electromagnetic field do not violate any symmetry and therefore are to be expected in an effective field theory with coupling constants inversely proportional to the Planck mass. Therefore, the leading term is likely to be linear in $\phi$.
At the same time the scalar field and its potential $V(\phi)$  also sets the value of the dark energy equation of state $w$.  As shown in Thompson (2012) this provides the link between the constants and the equation of state as
\begin{equation}
(1+w) = \frac{(\mu^\prime/\mu)^2}{3\zeta_\mu^2\Omega_\phi} = \frac{(\alpha^\prime/\alpha)^2}{3\zeta_\alpha^2\Omega_\phi} ~,
\end{equation}

where $\prime$ refers to the derivative with respect to $\log a$, where $a$ is the scale factor, 
$\Omega_\phi$ is the ratio of the dark energy density to the critical density and $\zeta_x$, $x$ = $\alpha$, $\mu$, is the coupling strength. The evolution of the fundamental constants is therefore driven by the equation of state.  Note that $w$ itself does not have to vary for $\mu^\prime$ or $\alpha^\prime$ to be non-zero;  $w$ just has to be different  from the $\Lambda$CDM value of -1.  The evolution of the fundamental constants can be calculated for various cosmologies, given their equation of state, or inversely the equation of state can be calculated from the observed constraints on the evolution of the fundamental constants.  These constraints produce regions in the $\zeta$, $(w+1)$  plane that are allowed and forbidden. In \cite{Thompson:2012pja,Thompson:2013xda} this landscape is examined for several cosmologies.  The conclusion is that either $w$ is within $0.001$ of $-1$ since $z=0.89$ (roughly half the age of the universe) or $\zeta_\mu$ is significantly below the lowest expected value, $\zeta_\mu=-4\times10^{-6}$ .  Figure \ref{constants1}  shows the current $\zeta$, $(w+1)$  landscape.

\subsection{Prospects for Research in Fundamental Constants}

It is clear that research in fundamental constants spans a broad range of areas in high energy physics, and astrophysics.  It also draws on the astronomical observational infrastructure for much of its data. Looking forward, there are various fruitful avenues to pursue.

Supersymmetry includes the possibility of a massless scalar particle sometimes called the dilaton.   The gauge coupling constants are set by the vacuum expectation value (VEV) of the dilaton.  A change in the VEV of the dilaton therefore induces changes in the value of both $\mu$ and $\alpha$.  Limits on the variation of these constants therefore provides a direct constraint on the variation of the dilaton VEV. As shown in Figure \ref{constants1} fundamental constant observations are already greatly limiting the expected value of the coupling.  Improved constraints expected in the next decade will limit this even more or set its value if changes in the fundamental constants are found and confirmed.

At this time very little is known about the true values or range of values for the coupling constants $\zeta_x$, or about how cosmological scalar fields couple to electromagnetic field. If we knew the value of $\zeta$ for each constant then the knowledge of the values of the constants as a function of redshift would immediately determine the dark energy equation of state as a function of redshift.  The value of $\zeta$, however, depends on the nature of the coupling between the cosmological or dilaton field and the electromagnetic field.  Currently most treatments consider a linear coupling. Other forms of the coupling may exist, such as a quadratic coupling, as suggested in~\cite{Olive:2001vz}, where $\phi$ couples to the electromagnetic kinetic term $F_{\mu\nu}F^{\mu\nu}$ as $(\phi - \phi_e)^2$ , where $\phi_e$ is some extremal value. Further research on the nature and range of the couplings is an essential part of addressing SUSY, the dark energy equation of state and cosmology.   
The predictions for the changes in the fundamental constants for several cosmologies, such as slow roll quintessence and $K$-essence, have been addressed in~\cite{Thompson:2012my,Thompson:2012pja,Thompson:2013xda} and described briefly here.  As more innovative ways to explain the accelerating universe continue to emerge, predictions on whether and how the fundamental constants would vary can become a useful check on these cosmologies.


In about two years the PEPSI optical spectrometer is expected to be operational on the Large Binocular Telescope (LBT) with a spectral resolution of 300,000.  This will increase the possible constraints on a change in $\mu$ by a factor of 50 at redshifts up to 4.  These observations will extend the current  radio constraints at redshifts below 1 to the high redshift regime to constrain the couplings and equation of state over 90\% of the age of the universe.

The next decade will see the start of observations with the LSST.  There is significant interest by the physics community in LSST as a mechanism for determining the dark energy equation of state.  In a 2012 Whitepaper  \cite{Albrecht:2009ct}, it is estimated that the linearized dark energy equation of state  can be determined at low redshift to an accuracy of  $\pm 0.05$  for $w_0$ and $\pm 0.15$  in $w_a$  at the 1$\sigma$ level.  Of course any finding of a non-zero value of  $w_a$   or $w_0 \neq -1$  would indicate significant new physics.  In terms of the fundamental constants, fixing the value of $(w+1)$  would set the acceptable range of the coupling constants $\zeta$ and the new physics they imply.  Conversely, added knowledge on couplings will add to the information gathered with LSST. 
There is also the possibility to study these fields with EUCLID and the SKA and its precursor MeerKAT.

\section{Future prospects for tests of gravity}

We have described  theories of modified gravity and dark sector couplings, and a new arena for tests of fundamental physics using astronomical phenomena and local experiments. Much of the work described above began in the last decade. So the story is rapidly evolving, with many  prospects for the next decade but, as of now, not a coherent observational program.
The exciting new realization is that we know enough about MG theories to design tests that are sensitive to broad classes of theoretical models, which make qualitatively similar predictions. We know the physical effects that can be detected via astronomical and local tests of gravity and even the scale for a signal. The decades old Parameterized-Post-Newtonian (PPN) program of null tests of GR can thus be enlarged into new arenas and connected to MG and other new physics. We describe a variety of low-budget  experiments below and discuss the kind of resources that would catalyze progress. Let's first look at the prospects for new theoretical models. 

\subsection{Theory}

An interesting and promising avenue for making theoretical progress is to better understand the properties of quantum field theories which rely on the Vainshtein mechanism to hide their additional degrees of freedom---in particular {\it massive gravity} and the closely related {\it galileon} theories. The higher-derivative structure of these theories necessitates a reorganization of standard effective field theory, indicating that they might not be best understood as standard Wilsonian effective theories.

In both massive gravity and the galileon---and more generally in theories which rely on Vainshtein screening---the recovery of Einstein gravity in the solar system manifestly relies on higher-derivative terms becoming important. Consider the simplest such interaction, $\sim\frac{1}{\Lambda^3}\partial^2\phi(\partial\phi)$. About a nontrivial background, $\phi=\bar\phi+\varphi$, this interaction term gives a contribution to the kinetic term $\sim \frac{1}{\Lambda^3}\partial^2\bar\phi(\partial\varphi)^2$; if we define $Z^2 \equiv \partial\bar\phi/\Lambda^3$, the Vainshtein mechanism relies on $Z$ being large. Canonically normalizing the fluctuation $\varphi$, we see that the cubic term about this background is actually suppressed by a much higher scale $\sim \frac{1}{Z\Lambda^3}\partial^2\varphi(\partial\varphi)^2$, one which is apparently above the cutoff of the original effective theory, $\Lambda$. This of course comes into conflict with our notions of how an effective field theory should work. In~\cite{Nicolis:2004qq}, it was shown that it is possible---under some optimistic assumptions---to make sense of these theories in the standard framework, but this certainly merits further study. A perhaps more radical alternative is that these theories should be understood not as Wilsonian theories in the normal sense: that is, theories that arise from some ultraviolet theory from integrating out some degrees of freedom. A reasonable question to ask is whether some sort of {\it quantum Vainshtein} effect also exists, leading to a re-summation of perturbation theory. One possible realization of thinking of this sort is so-called {\it classicalization} (see for example \cite{Dvali:2010jz, Dvali:2010ns}). Moving forward, understanding these novel quantum field theories should be an extremely interesting endeavor, regardless of the outcome for the purposes of cosmic acceleration.

There are interesting theoretical prospects for chameleon theories as well. The original motivation for chameleon models is to explain the moduli stabilization problem inherent in string and supergravity theories. Generically, the low energy effective theories include exactly massless fields - moduli - that should couple to all matter types with gravitational strength and thus induce unacceptable violations of known physics from big bang nucleosynthesis to solar system tests of gravity. Chameleon models provide a phenomenological solution to this problem but an exact top-down derivation from a fundamental theory is still required. Progress in this direction has been made already through a KKLT like approach, see \cite{Nastase:2013ik,Hinterbichler:2013we,Hinterbichler:2010wu}.
While the progress to date is good, there is still a need for a full UV completion of chameleon scalar fields. Indeed \cite{Erickcek} showed that a UV complete theory is essential to determine whether or not chameleon gravity is compatible with Big Bang Nucleosynthesis. 

In some respects screening mechanisms in modified gravity are victims of their own success---often times the bounds placed on the theories from solar system tests ensure that screening is so effective that the scalars are hidden from all current generation probes. It is therefore worthwhile to invest some effort into investigating novel setups where screening effects may be smaller. Some examples of this line of thinking for astrophysical probes of chameleon and symmetron screening are given in Sec. 1.5.3, but it is crucial for the observational and theoretical communities to communicate in order to find tests which are both interesting and possible. Another issue to address is the relative dearth of observational tests of the Vainshtein screening mechanism. Two promising areas to search to rectify this situation are to consider compact objects, which carry no scalar charge, and are therefore possibly unscreened~\cite{Hui:2012qt,Hui:2012jb} and to consider time-dependent situations, where the Vainshtein mechanism has been found to be less efficient than around static sources~\cite{deRham:2012fw,Chu:2012kz, deRham:2012fg}.

\subsection{Laboratory and solar system experiments}

Looking forward, there are already many planned experiments which will more strongly test gravity locally {\it i.e}, within the solar system and within the laboratory. The next-generation \eotwash~experiment will provide powerful constraints on gravitation-strength fifth forces at the dark energy scale, $2.4\times 10^{-3}$~eV$\sim 1/(80~\mu\textrm{m})$.  In particular, the simplest chameleon models will have large quantum corrections at laboratory densities unless their masses remain below $\sim 0.01$~eV, and \eotwash~will be able to exclude these quantum-stable models~\cite{Upadhye:2012qu}.  Constraints on larger matter couplings will come from planned bouncing neutron experiments such as GRANIT, which will improve by several orders of magnitude the measured energy splitting of neutrons in the Earth's gravitational field~\cite{Brax:2011hb}.  A proposed neutron interferometer at NIST will constrain the field gradients of matter coupled dark energy candidates, including chameleons and symmetrons, in a vacuum chamber~\cite{Schoen:2003my,Pokotilovski:2012js}.  For photon-coupled dark energy models, an adaptation of the CAST helioscope is planned which will constrain models with smaller matter couplings than afterglow experiments.

Following arguments presented in Section 3 of~\cite{Uzan:2010pm}, we can measure the limits on the variation of  $\alpha_{EM}$ with the ratio of frequencies of atomic clocks such as the ratio from rubidium and cesium clocks.  Here we use  $\alpha_{EM}$ to designate the value of $\alpha$ in electromagnetic interactions as opposed to nuclear interactions, holding open the possibility that they might be different.  In~\cite{Uzan:2010pm}, it is pointed out that these frequencies are proportional to the nucleon $g$ factors for each nucleus which depend on the proton and neutron $g$ factors.  These $g$ factors in turn depend on the light quark mass $m_q$  given by  $m_q = \frac{1}{2}(m_u + m_d)$ for the up and down quarks, the strange quark mass $m_s$ and the quantum chromodynamic scale.  The frequency of hyperfine transitions are proportional to a combinations of the g factors as well as  and $\mu$. This dependence means that variations in the fine structure constant $\alpha_{EM}$ and the proton to electron mass ratio $\mu$ imply variations in the quark masses as well.  Any detection of a variation in either $\mu$  or $\alpha$ then implies a variation in the quark masses with significant implications for high energy physics.  Limits on any such variation will be greatly improved by the advances in astronomical spectroscopy discussed above.

The current state of laboratory measurements of the fundamental constants is extensively reviewed in~\cite{Uzan:2010pm}.  If the accuracy of atomic clocks continues to improve at its current pace we would expect significant improvement in establishing the current value of the constants and placing limits on their time evolution to become competitive with the astrophysical measurements.

\begin{figure}[h]
\begin{center}
\includegraphics[width=4.5in]{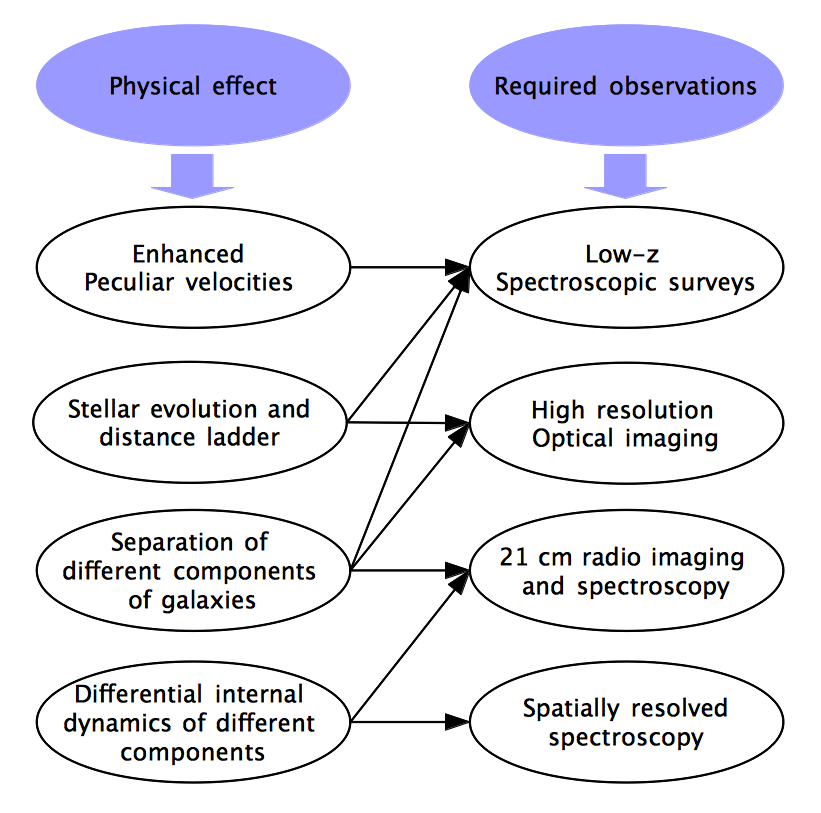}
\caption{Astrophysical tests of gravity described in  \ref{section:astrotests} mapped 
to the observations described in \ref{section:futureastro}. It is evident that more than one 
kind of telescope is needed for typical tests. However as discussed in the text, the sample 
size needed is modest and the galaxies are nearby, so they can be observed with high signal-to-noise using modest telescope resources. 
}
\label{figure:tests-observations}
\end{center}
\end{figure}

\subsection{Astrophysical tests}
\label{section:futureastro}

Recent work on astrophysical tests of gravity has shown that a new regime of tests exists using observations of stars, black holes and galaxies, see Section \ref{section:astrotests}. 
These small scale tests span a wide range of astrophysical environments and combine multi-wavelength data, often for a single test. However the 
sample size required for these tests is modest, typically hundreds of galaxies. So 
most of the promising tests can piggyback on planned surveys
or existing telescopes. 

The two key parameters of MG theories probed by astrophysical tests are the coupling of the 
scalar to matter and the range of the fifth force. 
In a large class of modified gravity theories including chameleons and symmetrons, fifth force source masses are screened according to their gravitational potentials.  
The program described for improving the bound on the self-screening parameter (related to the range of force) to $\sim 10^{-8}$, the smallest potentials accessible with dwarf galaxies, thereby eliminating the entire remaining parameter space of astrophysically testable models. For Vainshtein theories the astrophysical tests are less mature, but possibly even more promising since current limits come only from cosmological scales and are specific to the DGP model. A summary of the some of the astronomical resources needed for the tests described in Section  \ref{section:astrotests} follows. Figure \ref{figure:tests-observations} shows the mapping of physical effects to particular observational methods. The tables below list the tests according to the length scales probed.

\begin{itemize}
\item Low redshift spectroscopy:
Spectroscopic observations of samples of  
galaxies at low-$z$ is essential for a variety of tests of gravity.  
Such observations provide a detailed map of the nearby universe and can be used to extract 
the velocity field traced by galaxies. They can be carried out as part of
cosmological BAO surveys, and also by spectroscopic cameras on other telescopes. Multi-slit or fiber spectrographs on a number of telescopes are suitable.  

\item Spatially resolved spectroscopy. The internal dynamics of different tracers of galaxies (stars, ionized and neutral gas clouds) requires specialized observations. Some data exist on dwarf galaxies exists, whose principal purpose was to test for the nature of the inner dark matter profile  and those can be examined for differential motions. Current galaxy surveys using optical telescopes (SAURON, ATLAS-3D, CALIFA) focus on early type   or massive galaxies. A larger survey of low mass galaxies, for example pursued with the MANGA spectrographs of SDSSIV, could  increase the sample by more than an order of magnitude and address systematics by sampling environments that provide both screened and unscreened galaxies. 
Such observations may need
to be supplemented with 21cm radio observations to compare the stars and gas (see below). 

\item High redshift spectroscopy: Detailed spectroscopic observations on 8m class telescopes are needed for improvements in the tests of time variation of fundamental constants and coupled scalar fields.

\item High resolution imaging:
Imaging surveys require less modification for other gravity tests since targeting is not required: they
 cover large contiguous areas and a wide range
in redshift. Even so, samples of low-z galaxies will need to be observed with
higher resolution, feasible from space or adaptive optics telescopes on the ground.

Tests that utilize the distance ladder rely on HST observations of dwarf galaxies, in particular
for cepheid and TRGB distances.   This work would follow along the lines of the existing work (ground based observations for nearer galaxies, HST for those a bit further) and so does not require new capabilities. In the future, adaptive optics capabilities can extend the range over which some of these techniques (e.g. Cepheids) could be applied. Finally, wide field, highly sensitive narrow band imagers, such as the MMTF instrument on the Magellan telescope,  can effectively measure properties such as the planetary nebulae luminosity function in local dwarf galaxies. 

\item Radio observations:
Improved resolution observations at 21cm are needed
to test the MG predictions for the HI gas. The spatial resolution needed is up to an order of magnitude 
better than the recent ALFALFA survey, whose focus was on the velocity information. 
The radio observations
would be compared to optical data for the
stellar disks. Samples of order a hundred galaxies would suffice for useful tests and can 
be obtained  with instruments such as the eVLA. 
\end{itemize}

\begin{table*}[th]
\small
\begin{center}
\begin{tabular}{| m{2.7cm}| m{2.8cm}| m{2.5cm}| m{2.5cm}| m{4cm}|}
\hline
Test &   Theories Probed  & Current Status &  Prospects for next decade  & MG Signal             \\
\hline
Growth \- vs. \ \ Expansion: \ \ \ \ \ \ \ \ \ \ 100Mpc-1Gpc &
Test of GR + smooth dark energy &
$10$\% accuracy &
2-4\% accuracy by combining probes$^1$   &
MG signal is model-dependent \\
\hline
Lensing  vs. \ \ Dynamical mass$^2$: \ \ \ \ \ 0.01-100Mpc &
Test of GR &
20\% accuracy$^3$  &
5\% accuracy &
MG signal at 10\% level \\
\hline
Astrophysical Tests: \ \ \ \ \ \ \ \ \ \ \ \ \ \ \ 
0.01AU-1Mpc & MG  Screening Mechanisms & Few tests at the 10\% level & Several tests with factor of $10$ or more improvement & Chameleon and Vainshtein theories predict detectable signal  for several tests \\
\hline
Lab and Solar System Tests:  \ \ \ \ \ 1mm-1AU & PPN $\rightarrow$ MG parameters & Mass \& Range of 5th force$^4$ & Up to a factor of 10 improvement & Model-dependent$^5$ \\
\hline
\end{tabular}
\end{center}
\caption{Experimental tests of gravity and dark sector couplings, from mm to Gpc scales. A rough guide to the expected signal and experimental accuracy is given; see text for details. Footnotes: 1. BAO, SN, WL, Clusters, RSD, CMB lensing. 2. The test can be done over a range of scales: using strong lensing and stellar velocities inside galaxies, to cosmological scales using cross-correlations. 3. On scales where MG signatures are expected~\cite{Reyes:2010tr}. 4. See Section \ref{s:labtests}. 5. Also tests dark sector couplings.
\label{gravitytable}}
\end{table*}

Carrying out specific tests requires
 two additional elements: 
 
 \begin{itemize}
 \item
 A 3-dimensional map of the gravitational field in the  
 nearby universe is essential
 to quantify the screening levels in  chameleon and Vainshtein theories. Given a tracer of the 
 mass distribution, typically the optical light distribution coupled with information on the velocity field, one can determine the gravitational field in both GR and a MG theory. 
 This map making exercise for the nearby universe (out to 100s of Mpc) 
 is an essential part of the observational program.  It is feasible using currently planned wide 
 area surveys and additional spectroscopy (described above). 
 \item 
 Determining the level of screening for a MG model  requires solving the nonlinear equations of the theory.  Progress has been made in this challenging numerical exercise for the $f(R)$ and DGP models, and we expect future work will be needed to explore new models. This work involves 
 a collaboration between gravity theorists and numerical cosmologists. It will become more important 
 as detailed connections between specific observations and theories are made via numerical 
 realizations of the survey geometry. This work will also prepare us to figure out what a detection of new physics on small scales means for cosmological tests and vice versa. 
  \end{itemize}

\begin{table*}[h]
\small
\begin{center}
\begin{tabular}{| m{4cm}| m{5cm}| m{6cm}| }
\hline
Test &   Observations  & Telescopes and Facilities          \\
\hline
Consistency of growth and expansion& Planned dark energy program with imaging and spectroscopic surveys & DES, BOSS, DESI, LSST, Euclid, WFIRST \\
\hline
Lensing vs. Dynamical mass & Imaging and spectroscopic observations  & Fraction of time on planned surveys +\ $O$(100) nights on 4-m class spectroscopic telescopes\\
\hline
Astrophysical Tests &
Multi-wavelength observations of nearby galaxies &
2-4m class IFU spectroscopic telescopes; High resolution optical imaging; High resolution radio (21cm) observations\\
\hline
Lab and Solar System Tests &
Sensitivity improvements to torsion pendulum, neutron, and lunar laser ranging experiments; adaptations to axion experiments &
E\"ot-Wash, NIST, JLab, CAST \\
\hline
\end{tabular}
\end{center}
\caption{Resources needed for experimental tests of gravity in the next decade. See Figure \ref{figure:tests-observations} and Table \ref{gravitytable} for information on the tests. 
\label{futuretable}}
\end{table*}

\section{Conclusions}

A suite of novel probes of gravity and dark energy have demonstrated the capability to discover alternative explanations for the accelerating universe and new physics in the dark sector. These tests lie outside the regime of cosmological surveys but can piggyback on existing surveys and telescopes once careful feasibility studies are carried out.

The various tests described above have been carried out by independent teams, but the theoretical basis now exists to connect future experiments into a coherent program of testing
fundamental physics.  The investment needed is (a) to enable a modest augmentation of existing astronomical surveys, mainly by targeting new samples of galaxies, (b) to use  existing laboratory resources and telescopes, such as 2-4m class spectroscopic telescopes, to carry out new experiments or observations, and, (c) to promote an analysis program that features close coordination between theory, numerical simulations and data analysis.


%% file: NovelProbes/authorlist.tex


\begin{center}

\begin{large} {\bf Convener: Bhuvnesh Jain} \end{large}

Austin Joyce,
Rodger Thompson, 
Amol Upadhye, 
James Battat,
Philippe Brax, 
Anne-Christine Davis, 
Claudia de Rham, 
Scott Dodelson, 
Adrienne Erickcek,
Gregory Gabadadze, 
Wayne Hu, 
Lam Hui,
Dragan Huterer,
Marc Kamionkowski, 
Justin Khoury,
Kazuya Koyama, 
Baojiu Li, 
Eric Linder, 
Fabian Schmidt, 
Roman Scoccimarro, 
Glenn Starkman, 
Chris Stubbs,
Masahiro Takada, 
Andrew Tolley,
Mark Trodden,
Jean-Philippe Uzan, 
Vinu Vikram,
Amanda Weltman,
Mark Wyman, 
Dennis Zaritsky,
Gongbo Zhao
\end{center}
